\documentclass[reprint,5p,times,twocolumn]{elsarticle}

\usepackage[utf8]{inputenc}
\usepackage{amssymb}
\usepackage{xspace}
\usepackage{lineno}
\usepackage{graphicx}
\usepackage{amsmath}
\usepackage{xcolor}
\usepackage{natbib}

\usepackage{booktabs}
\usepackage{tabularx}

\usepackage[normalem]{ulem}

\usepackage[colorlinks,breaklinks,hidelinks]{hyperref}
\hypersetup{
  colorlinks = true
}

\usepackage[defaultcolor=blue]{changes}
\definechangesauthor[color=red]{AB}

\journal{Vacuum}

\begin{document}

\begin{frontmatter}

\title{Low stray-field ion pump configuration for cold atom experiments}

\author[a]{Preeti Sharma}
\affiliation[a]{organization={LP2N, Laboratoire Photonique, Numérique et Nanosciences, Université de Bordeaux-IOGS-CNRS},
            city={Talence},
            postcode={33400}, 
            country={France}}
\author[a]{Luisa Loranca Cruz}
\author[b]{Leonardo Ricci}
\affiliation[b]{organization={Department of Physics, University of Trento},
            city={Trento},
            postcode={38123}, 
            country={Italy}}
\author[a]{Andrea Bertoldi}
%\email{andrea.bertoldi@institutoptique.fr}
\begin{abstract}
Ion pumps used in cold-atom experiments generate stray magnetic fields that can perturb magneto-optical trapping, shift Zeeman sensitive energy levels, and introduce systematic errors in precision spectroscopy. We numerically investigate a compact low-stray-field ion pump configuration designed to reduce magnetic flux leakage through the ultrahigh-vacuum port without increasing the overall pump size. Starting from a simplified magnetic model of a standard ion pump, we study two complementary modifications: auxiliary permanent magnets surrounding the original pump magnets with opposite magnetisation, and a soft-magnetic honeycomb field stopper integrated at the pump port. The resulting combined configuration strongly reduces the magnetic flux density outside the pump. In the geometry considered here, the field at 10 mm from the UHV port is reduced by 97.9 \%, while the calculated reduction in effective pumping speed due to the honeycomb stopper is 17.9\% . These results indicate that engineered magnetic compensation and field-guiding structures can support more compact cold-atom vacuum systems while preserving pumping performance. Further work should validate the design with nonlinear magnetostatic modelling, molecular-flow simulations, and measurements on a complete pump geometry.

\end{abstract}


\end{frontmatter}

\section{Introduction}

Ion pumps are widely used to generate and maintain ultrahigh vacuum (UHV) in precision experiments, including cold-atom systems, high-resolution microscopy, particle accelerators, and space-compatible instruments \cite{Benvenuti2001,Poppa2004,Grangeon2004,Aguilera2014,Liu2021,Vovrosh2018}.
Their operation relies on a strong magnetic field to sustain the Penning discharge by forcing electrons onto long trajectories inside the anode cells \cite{Audi1987}. A drawback of this operating principle is the presence of stray magnetic fields outside the pump housing. Such leakage can become a limiting factor when the pump must be installed close to a magnetically sensitive experimental region.

This issue is particularly relevant in cold-atom experiments, where UHV and magnetic field control must be achieved simultaneously \cite{Aguilera2014,Liu2021,Ren2015,Takamoto2020,Hobson2022}. Pressures in the $10^{-9}$–$10^{-10}$ mbar range are typically required to reduce collisions between trapped atoms and background gas molecules, thereby preserving atom number, trap lifetime and coherence. At the same time, magneto-optical traps, magnetic transport stages, optical lattices, cavity-coupled systems, atomic clocks and atom interferometers require well-controlled magnetic environments, since uncontrolled bias fields and gradients can perturb cooling and trapping conditions, shift Zeeman-sensitive transitions, and contribute systematic errors in precision measurements \cite{Peters2001}. 

\begin{figure}
	\centering 
	\includegraphics[width=1\columnwidth]{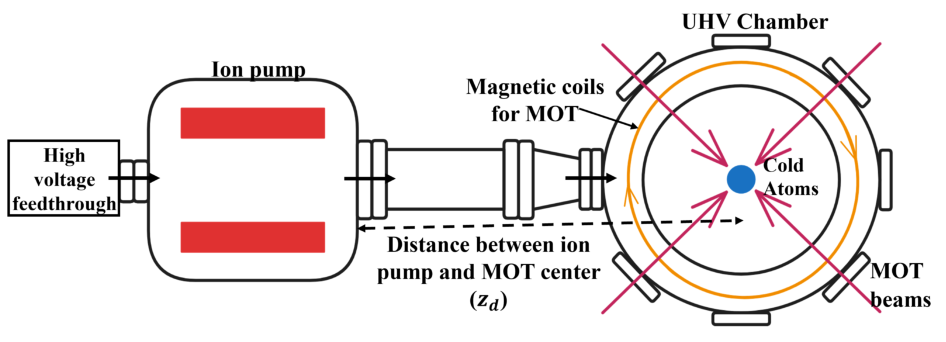}	
	\caption{Schematic of setup used for magneto-optical trap with ion pump kept at a distance ($z_{d}$) to minimise the effect of stray magnetic field generated by the pump.} 
	\label{fig:fig1n}
\end{figure}

A common practical solution is to place the ion pump far from the cold-atom region as shown in Figure \ref{fig:fig1n}. However, this approach increases the volume and conductance length of the vacuum system, which can require higher nominal pumping speeds, longer pump-down times, and more complex mechanical assemblies. These drawbacks are especially severe in compact sensors, transportable quantum devices, and space-based cold-atom experiments, where size, mass, power consumption, and mechanical simplicity are critical constraints \cite{Grangeon2004,Aguilera2014,Liu2021,Little2021,Ren2015}. Reducing the magnetic field leakage at the pump would enable more compact vacuum architectures while preserving the magnetic environment needed for atom manipulation. Non-evaporable getter (NEG) pumps provide an attractive alternative for magnetically sensitive UHV systems, as their operation does not require a magnetic field.
\begin{figure*}
	\centering 
	\includegraphics[width=1.9\columnwidth]{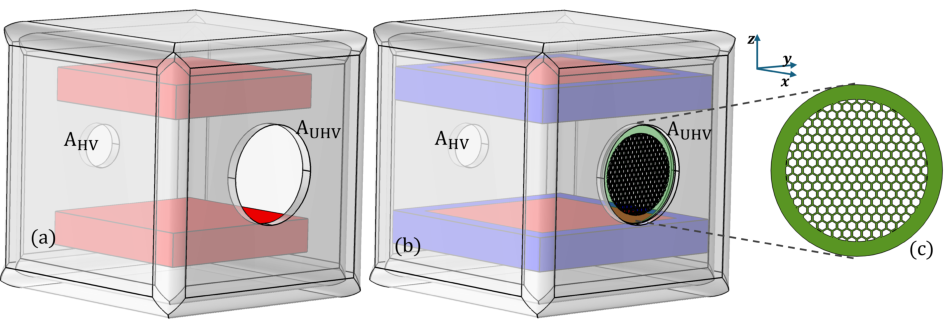}	
	\caption{(a) Standard configuration (SC) of the magnetic circuit of an ion pump: the gray thick-walled box represents the magnetic shield in soft-magnetic material, the two red parallelepipeds represent permanent magnets magnetised along the $+z$ direction, $\text{A}_{\text{HV}}$ denotes the aperture for the high voltage supply, and $\text{A}_{\text{UHV}}$ denotes the aperture towards the UHV chamber. (b) Two modifications are applied to the SC: the red parallelepipeds are surrounded by auxiliary magnets with opposite magnetisation (blue), and a stopper is installed at the $\text{A}_{\text{UHV}}$ port. When used independently, these modifications define Configurations I and II, respectively; when combined they define Configuration III. (c) Zoomed view of the stopper attached to $\text{A}_{\text{UHV}}$ on $xz$ plane.} 
	\label{fig:fig1}
\end{figure*}

In this work, we numerically study a low stray-field ion pump configuration based on two complementary design modifications. First, the magnetic return path is modified by adding auxiliary permanent magnets around the original pump magnets, with magnetisation direction opposite to that of the internal magnets. This arrangement redirects the magnetic flux and reduces the magnetic field escaping the pump body. Second, a soft-magnetic honeycomb field-stopper is introduced at the UHV port. This component mitigates the degradation of magnetic field containment caused by the aperture in the soft-magnetic shield, guiding residual magnetic flux while preserving vacuum conductance through a large number of small apertures. Together, these two elements reduce magnetic flux leakage through the vacuum port without increasing the overall pump size, and with only a modest calculated reduction in effective pumping speed. 

To make this comparison quantitative, we use two performance metrics. The first is the residual magnetic flux density, $\| \bf{B} \|$, along the central axis of the UHV port, evaluated at representative distances from the port. The second is the distance from the port at which the residual field falls below a specified application-dependent tolerance. For typical cold-atom operation, this tolerance depends on whether the relevant limitation is a static bias-field offset, a field gradient across the atom cloud, or a Zeeman shift in a precision measurement. In the present design study we therefore report the field at fixed distances and the inversion-point position, while treating the final acceptable field threshold as experiment-specific

This article is organised as follows. Section 2 describes the standard ion pump magnetic model and the proposed modified configurations. Section 3 estimates the conductance penalty introduced by the honeycomb stopper under molecular flow conditions. Section 4 presents the magnetostatic simulation model used to compare the standard and modified conditions. Section 5 presents the results, a parametric optimisation of the auxiliary-magnet size, and practical considerations for the implementation of the solutions in existing UHV systems. Finally, Section 6 discusses the expected impact on cold atom experiments.

\section{Ion pump configurations}

It must be noted that the standard configuration considered in this work is not intended to reproduce a specific commercial ion-pump model. Rather, it represents a simplified reference geometry used to isolate the effect of the proposed magnetic modifications. 
The main geometrical and magnetic parameters of the reference configuration are summarised in Table~\ref{tab:reference_pump}.

The magnetic model considered as a standard configuration (SC) consists of two identical permanent magnets placed inside a cubic-shell soft-magnetic return structure, as shown in Fig. \ref{fig:fig1}(a). The two magnets are separated along the $z$ direction and generate the magnetic field required for the Penning discharge. The surrounding soft-magnetic enclosure provides a low-reluctance return path for the magnetic flux and therefore limits the amount of magnetic field leaking outside the pump body. 
\begin{table}[t]
\centering
\small
\setlength{\tabcolsep}{4pt}
\caption{Main parameters of the reference ion-pump configuration used in the simulations.}
\label{tab:reference_pump}
\begin{tabular}{lll}
\hline
Parameter & Symbol & Value \\
\hline
Nominal pumping speed & $S_0$ & $20~\mathrm{L\,s^{-1}}$ \\
Outer shield side length & $L_{\mathrm{out}}$ & $120~\mathrm{mm}$ \\
Inner shield side length & $L_{\mathrm{in}}$ & $110~\mathrm{mm}$ \\
Shield wall thickness & $t_{\mathrm{sh}}$ & $5~\mathrm{mm}$ \\
Inner-corner radius & $R_{\mathrm{c}}$ & $10~\mathrm{mm}$ \\
Relative permeability & $\mu_r$ & $4000$ \\
HV aperture diameter & $D_{\mathrm{HV}}$ & $20~\mathrm{mm}$ \\
UHV aperture diameter & $D_{\mathrm{UHV}}$ & $44.5~\mathrm{mm}$ \\
Magnet dimensions & --- & $80\times80\times15~\mathrm{mm^3}$ \\
Magnet remanence & $B_r$ & $1~\mathrm{T}$ \\
Magnet separation & $d_{\mathrm{mag}}$ & $50~\mathrm{mm}$ \\
Vacuum pipe diameter & $D_{\mathrm{pipe}}$ & $38.1~\mathrm{mm}$ \\
\hline
\end{tabular}
\end{table}
The external side length of the soft-magnetic box is 120 mm, while the internal side is 110 mm, corresponding to a wall thickness of 5 mm. The inner corners of the box are rounded with a radius of 10 mm in order to improve magnetic-flux guidance and avoid sharp geometric discontinuities. The relative permeability of the soft-magnetic material is taken as $\mu_r=4000$. The box has two apertures: the first one, labeled A$_{\rm{HV}}$, has a diameter of 20 mm and represents the access for the high-voltage feedthrough of the ion pump electrodes; the second one, labeled A$_{\rm{UHV}}$, has a diameter of 44.5 mm and represents the UHV connection to the experimental chamber. This diameter is compatible with a standard pipe for a CF40 flange. Each permanent magnet has dimensions 80 × 80 × 15 mm$^3$, a remanent flux density of 1 T, and magnetisation along the vertical direction. The two magnets are separated by 50 mm, producing a nearly homogeneous magnetic field in the Penning discharge region, while the soft-magnetic box acts as the magnetic return path.

Although the soft-magnetic enclosure confines most of the flux, residual leakage remains because of the finite permeability of the shielding material and due to the apertures required for electrical and vacuum access. Leakage through A$_{\rm{UHV}}$ is particularly critical for cold atom experiments, since this port faces the experimental chamber. In conventional layouts, this issue is mitigated by increasing the distance between the pump and the atoms, at the cost of increasing vacuum-system volume and reducing the effective pumping speed at the chamber. We consider two modifications aimed at reducing the magnetic field leakage through A$_{\rm{UHV}}$, while preserving a compact vacuum layout.

In configuration I, auxiliary permanent magnets are added around the original magnets with respect to SC: each original magnet is laterally surrounded by a rectangular-ring magnet of the same thickness. The resulting composite magnet has a larger square footprint, with side length $(80+2x)$ mm, where $x$ is the added lateral size on each side used later in the optimisation. The auxiliary magnets have a remanent flux density of 1 T, but their magnetisation is oriented opposite to that of the original magnets. This compensation geometry modifies the global magnetic return path and reduces the flux escaping through the UHV port.

Configuration II introduces a magnetic field stopper at A$_{\rm{UHV}}$ with respect to SC. In the proposed implementation, the stopper is not a separate free-standing component, but a thin soft-magnetic honeycomb grid welded to the vacuum pipe that crosses the A$_{\rm{UHV}}$ aperture. Its role is to preserve vacuum conductance while partially closing the magnetic discontinuity created by the UHV aperture, thereby guiding stray flux back into the soft-magnetic structure. Candidate materials for the stopper include high-purity or low-carbon soft iron and high-permeability Fe--Ni alloys, which have been considered for magnetic shielding under UHV conditions \cite{Vovrosh2018,Kamiya2013}. The final choice requires a compromise between magnetic permeability, saturation magnetisation, UHV compatibility, and manufacturability. In particular, machining, forming, or welding can degrade the magnetic properties of high-permeability materials, so that an appropriate magnetic annealing procedure may be required after fabrication. The material should also exhibit sufficiently low outgassing and preserve its magnetic properties after the thermal cycles required for UHV bake-out. The present choice of $\mu_r=4000$ should therefore be regarded as a representative value for the proof-of-principle simulations rather than as the specification of a particular material. 

Configuration III adds both the auxiliary permanent magnets of configuration I and the magnetic field stopper of configuration II to SC, targeting a combined performance improvement. The three-dimensional rendering of the full assembly along with a zoomed view of the honeycomb stopper is shown in Figs. \ref{fig:fig1}(b) and (c), respectively.

When preserving the maximum pumping speed is the primary constraint, configuration I can be implemented without adding any element in the vacuum conductance path. Conversely, the honeycomb stopper directly reduces UHV-port leakage, with the conductance trade-off quantified in Sec.~\ref{Sec:vacuumConductance}. Combining both modifications provides the strongest attenuation of the stray field. This combined configuration is therefore the most suitable option for demanding cold-atom experiments in which the ion pump must be placed close to magnetically sensitive regions.

\section{\label{Sec:vacuumConductance}Vacuum conductance of the honeycomb stopper}

The honeycomb magnetic stopper is placed in the UHV pumping path and therefore introduces an additional conductance limitation. Since the system operates in the UHV regime, the relevant gas-flow regime is molecular flow. In this regime, the conductance of an aperture of finite thickness is not determined only by its open area but also by the molecular transmission probability through the aperture \cite{Carpenter1983}.

For a single channel, the molecular-flow conductance can be written as
\begin{equation}
C = \frac{1}{4} A \bar{v} W ,
\end{equation}
where \(A\) is the aperture area, \(\bar{v}\) is the mean molecular speed of the gas, and \(W\) is the Clausing transmission probability. For air at room temperature, this expression gives
\begin{equation}
C = 0.116\, A_{\mathrm{mm^2}}\, W
\quad \mathrm{L\,s^{-1}},
\end{equation}
with $A$ expressed in \(\mathrm{mm^2}\). 
The factor \(W=1\) corresponds to an ideal zero-thickness aperture. The stopper has a finite thickness of \(l = 1~\mathrm{mm}\), and the conductance must therefore be corrected by the transmission probability of the finite-length channels.

The holes of the honeycomb stopper are regular hexagonal apertures to optimise the packing efficiency and open-area fraction of the repeating lattice while preserving a constant 0.5 mm wall thickness between neighbouring holes. Here, the quoted aperture size \(F\) corresponds to the width across-flats of the hexagon. The area of one regular hexagonal aperture is therefore
\begin{equation}
A_{\mathrm{hex}} = \frac{\sqrt{3}}{2} F^2 ,
\end{equation}
which for \(F = 2~\mathrm{mm}\), gives $A_{\mathrm{hex}} = 3.46$ mm$^2$. 

The zero-thickness conductance of one such aperture would then be $C_{\mathrm{hex},0}
= 0.402~\mathrm{L\,s^{-1}}$. The honeycomb stopper is assumed to replace a short section of the existing pipe, of length \(L=1~\mathrm{mm}\).
For an initial analytical estimate, the hexagonal channel is mapped onto an equivalent circular channel using its area-to-perimeter ratio. For the present geometry, this gives \cite{Szwemin2002}
\begin{equation}
W = 0.672.
\end{equation}
Thus, the conductance of a single finite-thickness hexagonal channel is estimated as
\begin{equation}
C_{\mathrm{hex}}
= C_{\mathrm{hex},0} W
= 0.27~\mathrm{L\,s^{-1}}.
\end{equation}
For the repeating honeycomb geometry with 0.5 mm walls and a 38.1 mm clipping diameter, the effective number of apertures is \(N \simeq 210\); therefore, the total conductance of the honeycomb stopper is 
\begin{equation}
C_{\mathrm{stop}}
= N C_{\mathrm{hex}}
= 56.7~\mathrm{L\,s^{-1}}.
\end{equation}

To validate this analytical estimate, a Monte Carlo molecular-flow calculation was performed using Molflow+ \cite{Kersevan2019}. The numerical geometry reproduces the 38.1 mm-diameter, 1 mm thick honeycomb considered in the magnetic model, with 2 mm hexagonal apertures separated by 0.5 mm walls. In contrast to the analytical estimate, the finite circular geometry and the partially clipped apertures at its perimeter are explicitly included. As a reference, an unobstructed 1 mm long cylindrical section gives a molecular transmission probability W$_{\rm{pipe}}$=0.974, corresponding to C$_{\rm{pipe}}$=129 Ls$^{-1}$, in excellent agreement with the analytical estimate of approximately 128 Ls$^{-1}$. For the honeycomb stopper, Molflow+ gives an effective transmission probability W$_{\rm{MC}}$=0.637 and a total conductance C$_{\rm{stop}}^{\rm{MC}}$=53.7 Ls$^{-1}$. The latter differs by only approximately 5\% from the analytical estimate of 56.7 Ls$^{-1}$, supporting the validity of the simplified conductance model.

The nominal pumping speed of the ion-pump assembly is taken to be $S_0 = 20~\mathrm{L\,s^{-1}}$. This value is assumed to represent the effective pumping speed of the complete pump and its standard connection up to the CF40 flange, including the conductance limitation of the existing 38.1 mm inner diameter pipe.

The relevant comparison is therefore between the conductance of the removed pipe section $C_{\mathrm{pipe},L}$, equal to $\simeq 128~\mathrm{L\,s^{-1}}$ 
when considering the geometrical properties and Clausing transmission probability, and the conductance of the stopper $C_{\mathrm{stop}}$ occupying the same axial length. In the molecular-flow regime, the modified pumping speed can be written as
\begin{equation}
S_{\mathrm{mod}}
=
\left(
\frac{1}{S_0}
-
\frac{1}{C_{\mathrm{pipe},L}}
+
\frac{1}{C_{\mathrm{stop}}^{\mathrm{MC}}}
\right)^{-1}
= 16.4 ~\mathrm{L\,s^{-1}}.
\end{equation}
The effective pumping speed estimated using the Monte Carlo conductances gives a reduction of 17.9\% relative to the nominal 20 L$\text{s}^{-1}$ pumping speed, 
while the analytical model gives a closely comparable reduction of 16.4\%. 
For the representative 38.1 mm inner diameter connection considered here, this conductance penalty is comparable to the conductance gain obtained by shortening the pump-to-chamber connection by approximately 15 mm.

The present stopper is a representative design point and not an optimised solution. Increasing the across-flats size of the honeycomb apertures would increase the conductance and reduce the pumping-speed penalty, but would also reduce the amount of soft-magnetic material available to guide residual flux. Conversely, smaller apertures improve magnetic closure of the UHV opening at the expense of conductance. A full optimisation should therefore minimise a combined objective function including the residual stray field at the target atom position and the effective pumping-speed loss.

\section{Simulation model}
All magnetostatic simulations were performed using COMSOL Multiphysics\textsuperscript{\textregistered}, version 6.2 (COMSOL AB, Stockholm, Sweden), using the \emph{Magnetic fields, No current (mfnc)} module under stationary conditions. In this formulation, the magnetic field is written in terms of the magnetic scalar potential $V_m$ \cite[Ch.~5]{Jackson1999}:
\begin{equation}
    \mathbf{H}=-\nabla V_m ,
\end{equation}
with the constraint
\begin{equation}
    \nabla \cdot \mathbf{B}=0 .
\end{equation}
The computational domains include air, with ($\mu_{r}=1$), together with the box shield and honeycomb stopper, both made of soft low-carbon iron and modelled as linear material with relative permeability $\mu_{r}=4000$, thus neglecting non-linear saturation. The permanent magnets are described through a remanent-flux-density formulation
\begin{equation}
\bf B=\mu_0\mu_r \bf{H}+ \bf B_{r}
\end{equation}

with $\| \bf{B_r} \|=1$ T. The direction of $\bf{B_r}$ is set according to the magnetisation of the original and auxiliary magnets in each configuration.

The magnetic model was embedded in an air domain sufficiently larger than the pump geometry to minimise artificial boundary effects. Open-boundary conditions were applied at the outer boundary of the computational domain, and continuity of the normal component of (\textbf{B}) and of the tangential component of (\textbf{H}) was imposed at material interfaces by the finite-element formulation.

A user-controlled free tetrahedral mesh was used for the simulations. The mesh parameters were based on the default settings provided by COMSOL, with the additional constraint that the minimum element size should be smaller than the side length of the hexagonal apertures to adequately resolve the honeycomb geometry. The maximum element size was 30~mm, the maximum element growth rate was 1.5, the curvature factor was 0.6, and the resolution of narrow regions was 0.5. The influence of mesh refinement was assessed by repeating the simulations for several values of the minimum element size. The calculated magnetic field showed negligible variation for minimum element sizes below 0.1~mm; this value was therefore adopted for the simulations presented here. Further refinement increased the computational cost without producing a significant change in the calculated magnetic-field distribution. 

For each configuration, the magnetic flux density was evaluated both inside and around the pump. Particular attention was devoted to the stray magnetic field outside the pump body, along the central axis extending outward from the UHV port. The center of $A_{\rm{UHV}}$ was used as the reference origin for these scans, corresponding to $(x,y,z) = (160,100,100)$ mm in the simulation geometry. In the vector plots, the arrow lengths are logarithmically scaled to increase the range of visible field amplitude.
\begin{figure}[b!]
	\centering 
	\includegraphics[width=\columnwidth]{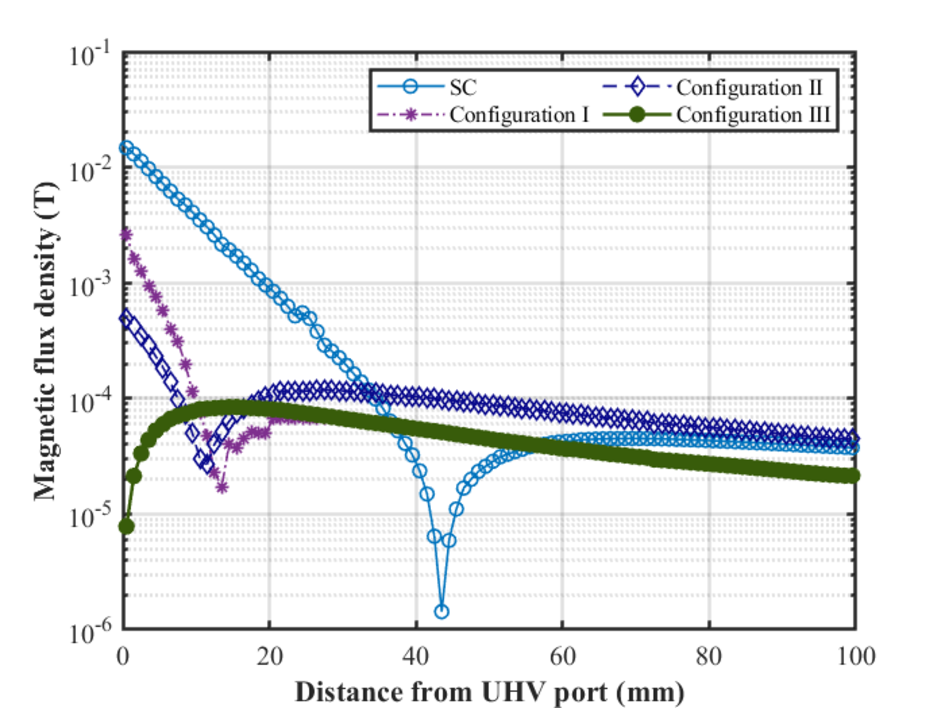}	
	\caption{Plot of magnetic flux density outside the box with respect to distance from $A_{\rm{UHV}}$ port at $(y,z)=(100,100)$ mm on semi-log scale for different configurations.}
	\label{fig:Final_allconfig_1Dplot}
\end{figure}
\section{Results and auxiliary-magnet size optimization}
\begin{figure*}[t!]
	\centering
	\includegraphics[width=2\columnwidth]{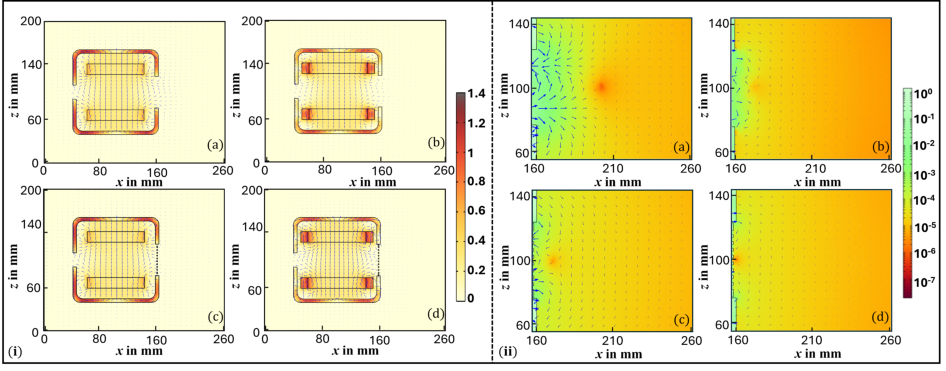}	
	\caption{Surface plot of magnetic flux density $(\vec{B})$ in the $xz$ plane (i) for: (a) SC, (b) Configuration I, (c) Configuration II, (d) Configuration III. The colour scale shows $|\vec{B}|$ in Tesla. Blue arrows indicate the projected magnetic field in the $xz$ plane; arrow lengths are  logarithmically scaled for visualisation only. (ii) Zoomed views of the region outside the UVH port using a logarithmic colour scale.} 
	\label{fig:allconfig}
\end{figure*}

The magnetic-field leakage was evaluated for the standard ion-pump geometry and for the three modified configurations introduced above. The magnetic flux density was first evaluated along the central axis extending outward from the UHV port, as shown in Fig.~\ref{fig:Final_allconfig_1Dplot}. This direction is the most relevant one for cold-atom experiments, because it corresponds to the line connecting the ion pump to the experimental chamber. In the SC case, the field decreases with distance from the port but remains significant over several centimeters. The field also exhibits a zero-crossing, or inversion point, relatively far from the pump aperture. This behaviour reflects the magnetic flux leaking through the UHV opening of the soft-magnetic return structure.

Adding the honeycomb magnetic stopper alone (at $x$ = 157 mm in our simulation), Configuration II, strongly reduces the magnetic field close to the UHV port. This confirms that the two mechanisms are complementary: the stopper mainly suppresses leakage through the UHV aperture, whereas the auxiliary magnets act on the global magnetic return path.

The strongest reduction is obtained for Configuration III, where the auxiliary magnets and the honeycomb stopper are used together. In this case, the field outside the UHV port is reduced both near the aperture and at larger distances. The position of the inversion point is also shifted substantially closer to the pump. In the simulations shown in Fig.~\ref{fig:Final_allconfig_1Dplot}, the inversion point is located at approximately \(43.5~\mathrm{mm}\) from the UHV port for the SC case, while it is reduced to about \(11.2~\mathrm{mm}\) when the stopper is added and to about \(0.5~\mathrm{mm}\) for the optimised combined configuration. These values are approximate, since they are extracted from field profiles calculated with a finite spatial step.

To provide an application-oriented quantitative criterion, we consider a stray magnetic-field threshold of 0.1~mT (1~G) at the position of the cold atoms. This value is not intended as a universal requirement, but is chosen to be of the same order of magnitude as the Earth's magnetic field, which generally has to be compensated or controlled in magnetically sensitive cold-atom experiments. Reducing the ion-pump contribution below this level therefore brings it into the range of environmental magnetic fields that can be addressed by the experiment's magnetic-field compensation system. From the axial profiles in Fig. \ref{fig:Final_allconfig_1Dplot}, the field of the standard configuration (SC) falls below 0.1~mT only at distances larger than approximately 35~mm from the UHV port, while exceeding 10~mT (100~G) close to the pump output. For Configuration I, the 0.1~mT threshold is reached at approximately 9~mm, with a field above 2~mT at the pump output. Configuration II suppresses the field directly at the aperture to approximately 0.5~mT, but the 0.1~mT threshold is satisfied only beyond approximately 45~mm. In contrast, Configuration III remains below 0.1~mT over the entire region outside the UHV port considered here. This comparison shows the advantage of combining the two complementary field-suppression mechanisms when the ion pump must be positioned close to a magnetically sensitive experimental region.

A quantitative comparison of the field reduction is summarised in Table~\ref{tab:field_configurations}. The table reports the magnetic flux density at selected distances from $A_{\rm{UHV}}$ and the corresponding reduction relative to the SC case. At a short distance from the port, the honeycomb stopper provides a particularly large reduction. At larger distances, the combined effect of the stopper and the auxiliary magnets remains the most efficient configuration. For experiments with a specified magnetic-field tolerance, the data in Table~\ref{tab:field_configurations} can be directly converted into a minimum pump-to-atom distance.
\begin{table*}[t]
\centering
\caption{Magnetic field values (in mT) for the different configurations. The reduction factor is calculated with respect to the standard configuration at the same distance.}
\label{tab:field_configurations}
\resizebox{\textwidth}{!}{
\begin{tabular}{lcccccccc}
\toprule
\textbf{Configuration}&
\textbf{\begin{tabular}{c}
B inversion point  \\
(in mm)
\end{tabular}} &
\textbf{\begin{tabular}{c}
$B_{max}$ at Penning  \\
discharge region
\end{tabular}} &
\textbf{\begin{tabular}{c}
10 mm \\
from $A_{\rm{UHV}}$
\end{tabular}} &
\textbf{\begin{tabular}{c}
30 mm \\
from $A_{\rm{UHV}}$
\end{tabular}} &
\textbf{\begin{tabular}{c}
50 mm \\
from $A_{\rm{UHV}}$
\end{tabular}} &
\textbf{\begin{tabular}{c}
100 mm \\
from $A_{\rm{UHV}}$
\end{tabular}} &
\textbf{\begin{tabular}{c}
Reduction vs SC \\
at 5 mm (in \%)
\end{tabular}} \\
\midrule
SC& 43.5 & 235 & 3.78 & 0.209 & 0.027 & 0.037 & 0 \\
Conf. I & 13.3 & 223 & 0.091 & 0.065 & 0.045 & 0.021 & 91.3 \\
Conf. II & 11.2 & 228 & 0.038 & 0.116 & 0.088 & 0.044 & 97.3 \\
Conf. III & 0.5 & 222 & 0.079 & 0.067 & 0.044 & 0.021 & 99.3 \\
\bottomrule
\end{tabular}%
}
\end{table*}
The spatial distribution of the magnetic field was then inspected using two-dimensional field maps in the \(xz\) plane, as shown in Fig.~\ref{fig:allconfig}. These maps confirm that the reduction is not limited to the central axis. In the SC case, magnetic flux lines escape through $A_{\rm{UHV}}$ and extend into the region where a vacuum chamber or cold-atom cell would be located. With the honeycomb stopper, the field is more efficiently guided within the soft-magnetic material and the leakage through the aperture is reduced. With the auxiliary magnets, the external field is further compensated by redistributing the flux inside the magnetic return structure. The combined configuration gives the smallest external field in the region facing the UHV port.

\begin{figure}[t!]
	\centering
	\includegraphics[width=\columnwidth]{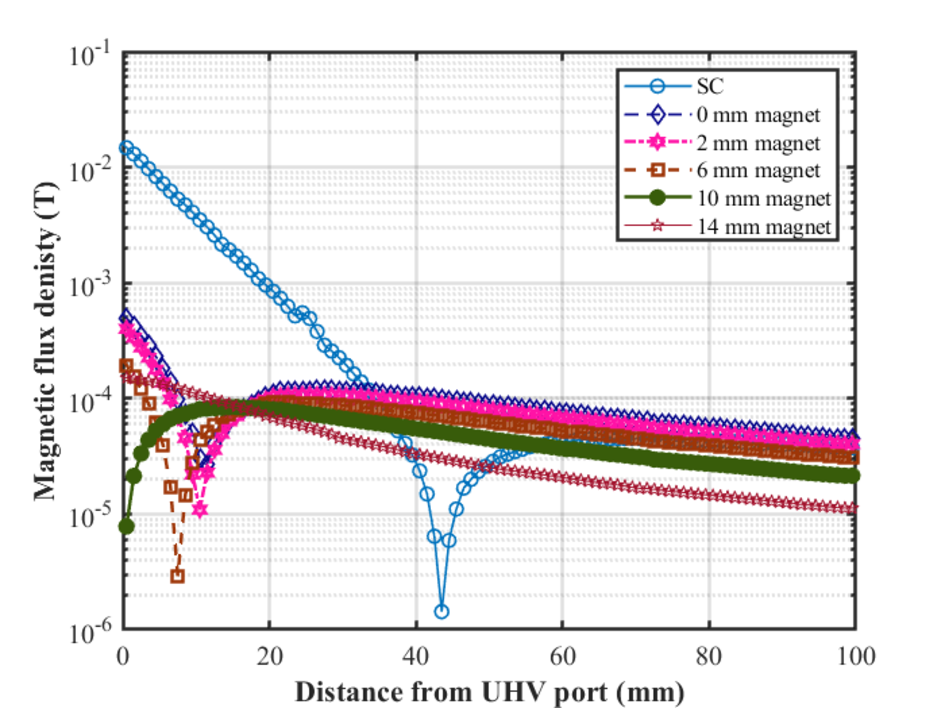}	
	\caption{Plot of magnetic flux density (in Tesla) outside the magnetic shield as a function of distance from the UHV port at $(y,z)=(100,100)$ mm on a semi-logarithmic scale for SC and configuration II with different sizes of auxiliary magnets.}
	\label{fig:magnetsize_1D}%
\end{figure}

A one-parameter optimisation of the auxiliary-magnet size was then performed for the combined configuration. The auxiliary magnets were implemented as rectangular-ring magnets surrounding the original permanent magnets, with magnetisation opposite to that of the original magnets. The added lateral size was varied between \(2~\mathrm{mm}\) and \(14~\mathrm{mm}\). The corresponding axial field profiles are shown in Fig.~\ref{fig:magnetsize_1D}. Increasing the auxiliary-magnet size reduces the stray field and shifts the inversion point closer to the pump. However, the improvement becomes marginal for added sizes larger than about \(10~\mathrm{mm}\). In addition, larger auxiliary magnets increase mechanical congestion inside the soft-magnetic enclosure and reduce the available clearance between the magnets and the shield.

On this basis, an added lateral size of \(10~\mathrm{mm}\) was selected as practical optimum within this one-parameter scan. This value provides a strong reduction of the external field while avoiding unnecessary increase in the magnet volume and mechanical complexity. For this selected geometry, the inversion point is located close to the UHV port, at approximately \(0.5~\mathrm{mm}\), and increasing the magnet size further does not produce a significant additional improvement in the simulated field profile.

An important constraint is that the modifications should not significantly degrade the magnetic field required for ion-pump operation. The auxiliary magnets are magnetised opposite to the original pump magnets and therefore must be designed so that they reduce the external stray field without suppressing the internal field in the Penning discharge region. The selected geometry was therefore checked to ensure that the field inside the pump remains compatible with ion-pump operation (as shown by $B_{max}$ in the third column of Table~\ref{tab:field_configurations}). The magnetic simulation establishes that the modification perturbs the characteristic Penning-region field by only ~5\%; determining the corresponding pumping-speed change requires a specific pump/electrode model or experimental characterisation. More generally, a pump-specific optimisation could include both the external stray field and the magnitude and spatial homogeneity of the magnetic field in the Penning-discharge region as simultaneous optimisation objectives. The auxiliary magnets could therefore potentially be designed to preserve, or even improve, the internal field distribution while minimising magnetic leakage.

Overall, Configuration III provides the strongest attenuation over the region facing the UHV port while preserving a compact pump geometry. Its practical use must therefore be assessed together with the conductance analysis of Sec.~\ref{Sec:vacuumConductance}.

\section{Discussion and outlook}

The simulations show that the combined configuration provides the strongest stray-field attenuation among the geometries considered. The improvement is obtained without increasing the overall pump size, but at the cost of design trade-offs involving vacuum conductance, magnetic-material properties, manufacturability, and UHV integration.

This result is relevant for cold-atom experiments, where ion pumps are often placed away from the science chamber to avoid magnetic perturbations. Reduced magnetic leakage could enable shorter pump-to-chamber connections, simplifying the vacuum system and improving the effective pumping speed at the science chamber. The main trade-off is the stopper conductance: for the representative geometry analysed here, the effective pumping speed decreases from 20 L s$^{-1}$ to 16.4 L s$^{-1}$, corresponding to a 17.9\% reduction. This penalty may be acceptable when magnetic-field suppression is the dominant constraint.

Several practical points must be considered before implementing the design in an experimental system. First, the stopper material must be compatible with UHV operation. It should have low outgassing, tolerate bake-out, and preserve its magnetic properties after thermal cycling. Second, the honeycomb geometry must be manufactured with sufficient precision to maintain a high open-area fraction while preserving mechanical rigidity. Third, the finite wall thickness, edge geometry, and possible partial blockage near the outer perimeter of the stopper should be included in a more detailed molecular-flow calculation. Fourth, the magnetic material should be checked for saturation. The present simulations assume a prescribed relative permeability for the soft-magnetic parts. If the stopper or shield approaches magnetic saturation, the actual field reduction may be smaller than predicted by the linear model. Other compensation geometries may be considered: for example, auxiliary magnets placed only on the side facing the UHV port, and maybe with magnetisation direction different from the z-axis would be mechanically simpler than the ring-shaped geometry studied here, and would modify the magnetic field only in the key target region.

The simulations use a simplified ion-pump geometry where the reference configuration was deliberately defined as a generic magnetic circuit to isolate the effects of the proposed modifications from manufacturer-specific design features. A real pump includes electrodes, anode structures, feedthroughs, welds, and material discontinuities that can affect both magnetic leakage and vacuum conductance. Moreover, the relation between the magnetic field in the Penning-discharge region and the actual pumping performance depends on the specific pump geometry and operating conditions. The present results should therefore be viewed as a proof-of-principle design study rather than as a performance prediction for a particular commercial pump. Application to an actual device will require validation using its complete geometry and manufacturer specifications, followed by experimental characterisation of the pumping performance and residual stray magnetic field.

Despite these limitations, the simulations identify a compact strategy for reducing ion-pump stray fields without increasing the overall pump size. In the simplified geometry considered here, the selected combined configuration reduces the magnetic flux density in the region facing the UHV port, shifts the magnetic-field inversion point close to the pump, and preserves a Penning-region field comparable to that of the standard configuration. These features support compact vacuum architectures with reduced conductance losses between the pump and the science chamber.

The final design must balance magnetic attenuation, effective pumping speed, UHV compatibility, magnetic-material saturation, manufacturability, and mechanical integration. Future work should combine Monte Carlo molecular-flow simulations of the honeycomb conductance, nonlinear magnetostatic simulations using the full pump geometry, and experimental validation of both the stray-field reduction and the effective pumping speed. These results should then support a multi-objective optimisation of the honeycomb aperture size, wall thickness, stopper thickness, and auxiliary-magnet geometry, with residual magnetic leakage and molecular-flow conductance used as the main optimisation targets.

\section*{Acknowledgements}
PS thanks AtomQTRL program for financial support. This work is partly supported by the European Union’s Horizon 2020 research and innovation programme (FET-Open project CRYST$^3$ N. 964531, project Qu-Test N. 101113901), Euramet (Project 23FUN02 CoCoRICO), Conseil Régional de Nouvelle Aquitaine and Naquidis Center (QPLEX project), the EUR Light S\&T Graduate Program (PIA3 Program “Investment for the Future”, ANR-17-EURE-0027), and the Idex of the University of Bordeaux (Research Program "GPR Light"). Qu-Test Project has received funding from the European Union’s Horizon Europe—The EU research and innovation program under the Grant Agreement 101113901.

\bibliographystyle{elsarticle-num} 
\bibliography{ref}

\end{document}